# Mechanical analogy for the wave-particle: helix on a vortex filament

Valery P. Dmitriyev

*Lomonosov University*
*P.O.Box 160 Moscow 117574 Russia*
*e-mail: dmitr@cc.nifhi.ac.ru*

The small amplitude-to-thread ratio helical configuration of a vortex filament in the ideal fluid behaves exactly as de Broglie wave. The complex-valued algebra of quantum mechanics finds a simple mechanical interpretation in terms of differential geometry of the space curve. The wave function takes the meaning of the velocity, with which the helix rotates about the screw axis. The helices differ in type of the screw - right or left-handed. Two kinds of the helical waves deflect in the inhomogeneous fluid vorticity field in the same way as spin particles in the Stern-Gerlach experiment.

The helix represents the low curvature asymptotics of a loop-shaped soliton, the latter being governed by the nonlinear Schroedinger equation. The length of the redundant segment, needed in order to form a curvilinear configuration on the originally straight vortex filament, measures the mass of a particle. The unique size of the loop on the vortex filament can be determined by the balance between the energy of the redundant segment and the energy due to the curvature of the loop.

The translational velocity of the soliton has the maximum at a value, which is inversely proportional to the length of the redundant segment. Insofar as the maximal velocity of a soliton is restricted from the above by the speed of the perturbation wave in the turbulent medium (i.e. the speed of light in vacuum), there must be a minimal redundant segment. Its length correlates with the Planck's constant.

In the stochastic environs a loop-shaped soliton disintegrates into the collection of the elementary asymptotic helices. An asymptotic helix obeys the linear Schroedinger equation with no dependence on mass. The mass of the particle appears explicitly when we describe the motion of the whole ensemble of the elementary splinters.

## 1. Introduction

Below, the earlier suggested [1] mechanical analogy for quantum particle is further developed. A helical wave on a vortex filament in the ideal fluid is considered. It is shown to obey the linear Schroedinger equation. Other properties of a vortex filament also reproduce the specific features of a quantum object.

This work is a constituent of the whole project aimed at constructing a regular mechanical analogy of physical fields and particles. The approach is based on the concept of a substratum for physics. The substratum is a universal medium serving to model the waves and action-at-a-distance in vacuum. This medium is viewed mesoscopically as a turbulent ideal fluid. Perturbations of the turbulence model physical fields. In this way the equations were derived that reproduce exactly [2] the Maxwell's electromagnetic equations. The voids in the fluid give rise to dilatational inclusions, which serve as a model [3, 4] of charged particles. Microscopically the turbulent substratum is seen as the vortex sponge. The latter is postulated as an ideal fluid pierced in all directions by the straight vortex tubes [5]. The hollow vortex tubes will be treated further as vortex filaments. We will consider a one-dimensional model of the vortex sponge with some recourse to higher dimensions. The microscopic construction presented here agrees well with respective mesoscopic models.

## 2. Vortex filament

The motion of an isolated vortex filament is governed by a dependence of the velocity $\mathbf{u}$ of the vortex filament's liquid element on the local form of the curve. To express such a law analytically one needs to describe the vortex filament as a space curve in the usual Frenet-Serret frame.

First, a point on a spatial curve is defined by the position vector $\mathbf{r}$, which is a function $\mathbf{r}(l)$ of the length $l$ measured from a fiducial point along the curve. For a moving curve, there is a further dependence $\mathbf{r}(l,t)$ on the time $t$. Excluding information about the curve's space position, the local form of the curve is fully specified by its curvature $\kappa(l,t)$ and torsion $\tau(l,t)$. The latter are defined through the two unit vectors, a tangent

$$\mathbf{e}(l,t) = \frac{\partial \mathbf{r}}{\partial l} \qquad (2.1)$$

and principal normal

$$\mathbf{n}(l,t)$$

(Fig.1), by the Frenet-Serret formulae

$$\kappa \mathbf{n} = \frac{\partial \mathbf{e}}{\partial l} \qquad (2.2)$$

$$\tau \mathbf{n} = -\frac{\partial (\mathbf{e} \times \mathbf{n})}{\partial l} \qquad (2.3)$$

$$|\mathbf{e}| = 1, \qquad |\mathbf{n}| = 1$$

The motion of the vortex filament without stretching is described in these terms by the Arms' equation

$$\mathbf{u}(l,t) = \frac{\partial \mathbf{r}}{\partial t} = \nu \, \kappa \, \mathbf{e} \times \mathbf{n} \qquad (2.4)$$

where $\nu$ stands for the coefficient of local self-induction and $\mathbf{e}$ is assumed to be parallel to the filament's vorticity vector (Fig.1; for a rigorous derivation see [6]). Using (2.1), (2.2), equation (2.4) can be rewritten in the straightforward form

$$\frac{\partial \mathbf{r}}{\partial t} = \nu \frac{\partial \mathbf{r}}{\partial l} \times \frac{\partial^2 \mathbf{r}}{\partial l^2} \qquad (2.5)$$

## 3. Small disturbances

Let the filament be directed along the $x$ axis. We will seek a solution to (2.5) in the form $\mathbf{r}(x,t)$ as small disturbances of the rectilinear configuration. That implies

$$\left|\frac{\partial y}{\partial x}\right|, \left|\frac{\partial z}{\partial x}\right|, \left|\frac{\partial^2 y}{\partial x^2}\right|, \left|\frac{\partial^2 z}{\partial x^2}\right| \ll 1$$

On this account, the corresponding quadratic terms will be further neglected throughout. So, we have for the arc's element

$$dl = \left[1 + \left(\frac{\partial y}{\partial x}\right)^2 + \left(\frac{\partial z}{\partial x}\right)^2\right]^{1/2} dx \approx dx$$

and (2.5) can be rewritten as

$$\frac{\partial \mathbf{r}}{\partial t} = \nu \frac{\partial \mathbf{r}}{\partial x} \times \frac{\partial^2 \mathbf{r}}{\partial x^2} \qquad (3.1)$$

We have

$$\mathbf{r}(x,t) = x\mathbf{i}_1 + y(x,t)\mathbf{i}_2 + z(x,t)\mathbf{i}_3$$

$$\frac{\partial \mathbf{r}}{\partial x} = \mathbf{i}_1 + \frac{\partial y}{\partial x}\mathbf{i}_2 + \frac{\partial z}{\partial x}\mathbf{i}_3$$

$$\frac{\partial^2 \mathbf{r}}{\partial x^2} = \frac{\partial^2 y}{\partial x^2}\mathbf{i}_2 + \frac{\partial^2 z}{\partial x^2}\mathbf{i}_3$$

That gives for (3.1)

$$\frac{\partial \mathbf{r}}{\partial t} = v\,\mathbf{i}_1 \times \left(\frac{\partial^2 y}{\partial x^2}\mathbf{i}_2 + \frac{\partial^2 z}{\partial x^2}\mathbf{i}_3\right) \tag{3.2}$$

Insofar as

$$\mathbf{i}_1 \times (y\mathbf{i}_2 + z\mathbf{i}_3) = -z\mathbf{i}_2 + y\mathbf{i}_3 \tag{3.3}$$

the right-hand side of (3.2) does not contain the $\mathbf{i}_1$ component. That enables us to drop the respective term in the left-hand side:

$$\frac{\partial y}{\partial t}\mathbf{i}_2 + \frac{\partial z}{\partial t}\mathbf{i}_3 = v\,\mathbf{i}_1 \times \left(\frac{\partial^2 y}{\partial x^2}\mathbf{i}_2 + \frac{\partial^2 z}{\partial x^2}\mathbf{i}_3\right) \tag{3.4}$$

The latter form is convenient for further applications. So, it can be taken as the basic equation for small disturbances of the vortex filament in the ideal fluid.

The simplest shape for the initial configuration of the filament is given by a curve with constant curvature $\kappa$ and $\tau$ torsion. Below, it will be treated in two representations, which are equivalent to each other. First, we will discuss it in vector form as implied by the equation (3.4).

## 4. Vector mechanics

We consider a right-hand screw helix positioned along the $x$ axis:

$$\begin{aligned} y &= a\cos(x/b) \\ z &= a\sin(x/b) \end{aligned} \tag{4.1}$$

where $a > 0$ is the amplitude and $b > 0$ the thread (or pitch) of the helix (Fig.2). This curve can be suggested as a small perturbation of the straight line if we take

$$a \ll b \tag{4.2}$$

That gives for (2.1)-(2.3) neglecting the small quantity $a^2/b^2$:

$$\mathbf{r} = x\,\mathbf{i}_1 + a\cos(x/b)\,\mathbf{i}_2 + a\sin(x/b)\,\mathbf{i}_3$$

$$dl = |d\mathbf{r}| = \left(1 + a^2/b^2\right)^{1/2} dx \approx dx$$

$$\mathbf{e} = \mathbf{i}_1 + \frac{a}{b}\left[-\sin(x/b)\,\mathbf{i}_2 + \cos(x/b)\,\mathbf{i}_3\right]$$

$$\kappa\mathbf{n} = -\frac{a}{b^2}\left[\cos(x/b)\,\mathbf{i}_2 + \sin(x/b)\,\mathbf{i}_3\right] \tag{4.3}$$

$$\mathbf{e}\times\mathbf{n} = (a/b)\,\mathbf{i}_1 + \sin(x/b)\,\mathbf{i}_2 - \cos(x/b)\,\mathbf{i}_3 \tag{4.4}$$

$$\tau\mathbf{n} = -\frac{\partial(\mathbf{e}\times\mathbf{n})}{\partial l} = -\frac{1}{b}\left[\cos(x/b)\,\mathbf{i}_2 + \sin(x/b)\,\mathbf{i}_3\right]$$

Therefrom, the curvature of the asymptotic helix is

$$\kappa = a/b^2 \tag{4.5}$$

and the torsion

$$\tau = 1/b \tag{4.6}$$

In these terms, the relation (4.2) looks as

$$\kappa \ll \tau \qquad (4.7)$$

It is implicit here that the direction of the filament's vorticity coincides with the vector $\mathbf{e}$. Hence, the motion of the filament can be calculated using (2.4). Substituting to it (4.4) with (4.5), (4.6) and neglecting the small velocity component along the $x$ axis, we get

$$\mathbf{u} = a v \tau^2 [\sin(\tau x) \mathbf{i}_2 - \cos(\tau x) \mathbf{i}_3] \qquad (4.8)$$

So, the helix rotates counterclockwise around the $x$ axis (looking along the $x$ axis) with the constant angular velocity

$$\omega = v \tau^2 \qquad (4.9)$$

With account of the angular displacement, the initial relation (4.1) should be improved

$$\begin{aligned} y &= a \cos(\tau x - v \tau^2 t) \\ z &= a \sin(\tau x - v \tau^2 t) \end{aligned} \qquad (4.10)$$

This provides the solution to the basic equation (3.4).

## 5. Schroedinger equation

You see from the above that when $a\tau \ll 1$ the motion of the vortex filament reduces itself to a plane vector mechanics. By virtue of this, relations (4.10) can be represented as a complex function $\varphi(x,t)$ of real variables:

$$\varphi(x,t) = a[\cos(\tau x - v\tau^2 t) + i \sin(\tau x - v\tau^2 t)] = a \exp[i(\tau x - v\tau^2 t)] \qquad (5.1)$$

In this connection, the vector form (3.3)

$$\mathbf{i}_1 \times (y \mathbf{i}_2 + z \mathbf{i}_3) = -z \mathbf{i}_2 + y \mathbf{i}_3$$

which the equation (3.4) is based on, corresponds to the relation for complex values

$$i(y + iz) = -z + iy$$

That puts (3.4) into the form of the Schroedinger equation

$$\frac{\partial \varphi}{\partial t} = i v \frac{\partial^2 \varphi}{\partial x^2} \qquad (5.2)$$

where

$$\varphi = y(x,t) + i z(x,t)$$

Equation (5.2), or (3.4), has the simple geometrical meaning. In a helix the principal normal $\mathbf{n}$ lays in a plane, which is perpendicular to the $x$ axis, and it is directed to the $x$ axis (see (4.3)). When $a\tau \ll 1$, the tangent $\mathbf{e}$ is almost parallel to the $x$ axis. So, in order to get from it the self-induction velocity (2.4), we must merely rotate $\mathbf{n}$ at the angle $\pi/2$ counterclockwise around the $x$ axis if looking against this axis. In terms of complex values, the curvature

$$\kappa n = \frac{\partial^2 \varphi}{\partial x^2}$$

The operation

$$i \kappa n$$

corresponds to the above mentioned rotation of $n$. The self-induction velocity is

$$u = \frac{\partial \varphi}{\partial t}$$

Here $\varphi$, $n$ and $u$ are complex values and $x$, $t$, $\kappa$ real values.

## 6. The wave packet

So, the asymptotic solution above discussed can be represented in the form of the hypercomplex value $r$:

$$r(x,t) = i_1 x + \varphi(x,t)$$

where $x$, $t$ are real values. According to the above found, the complex-valued function $\varphi(x,t)$ can be expanded into the sum of harmonics

$$\varphi(x,t) = \int c(\tau) \exp[i(\tau x - v\tau^2 t)] d\tau$$

As usual, taking this integral in the range $[\tau_o - \Delta\tau, \tau_0 + \Delta\tau]$, we get the wave packet

$$a \frac{\sin[(x - 2v\tau_0)\Delta\tau]}{(x - 2v\tau_0)\Delta\tau} \exp[i(\tau_0 x - v\tau_0^2 t)] \qquad (6.1)$$

(Fig.3). The hump of the wave packet moves translationally with the velocity

$$\upsilon = 2v\tau_0 \qquad (6.2)$$

The remarkable feature of this phenomenon is that the motion of the hump is owing to the rotation of an individual helix with the angular velocity $v\tau^2$ but not because of its longitudinal motion. This is the effect of a screw! A bolt is screwed into a nut due to rotation. In general, the velocity $\upsilon$ of screwing in depends on the thread $b$ as $\omega\, b$. From (4.9), (4.6), for the vortex helix $\omega \sim 1/b^2$. Therefore, $\upsilon \sim 1/b$, that is in accord with (6.2).

As you will see further, the wave packet gives us an approximation for asymptotic solution to nonlinear equation (2.5) constructed from the solutions to corresponding linear equation (3.4).

## 7. Soliton

The equation (2.5) possesses [7] the following exact solution $\mathbf{r}(l,t)$:

$$\mathbf{r} = x\mathbf{i}_1 + y\mathbf{i}_2 + z\mathbf{i}_3$$

$$x = l - a \tanh\eta \qquad (7.1)$$

$$y + iz = a\, \mathrm{sech}\eta\, \exp(i\theta)$$

where

$$a = 2\hat{\kappa}/(\hat{\kappa}^2 + \tau^2) \qquad (7.2)$$

$$\eta = \hat{\kappa}\left(l - 2v\,\tau\, t\right) \qquad (7.3)$$

$$\theta = \tau l + v(\hat{\kappa}^2 - \tau^2) t$$

$$\tau = \mathrm{const}$$

$$\hat{\kappa} = \mathrm{const}$$

As before, $v$ is the self-induction coefficient of the vortex filament.

In order to form the curvilinear configuration on the straight line, one needs an extra segment of the filament, which will be further referred to as the redundant segment. Its length is easily found integrating the differential of (7.1) all over the $x$ axis

$$\int dx = \int dl - a \int d\tanh\eta$$

whence

$$[l - x]_{-\infty}^{+\infty} = 2a \qquad (7.4)$$

Substituting $\mathbf{r}(l,t)$ to (2.2) we find that the curvature of the line is described by the bell-shaped function

$$\kappa(l,t) = 2\hat{\kappa}\,\text{sech}\,\eta \qquad (7.5)$$

Its parameter $1/\hat{\kappa}$ from (7.3) can be used as a measure of the disturbance's delocalization. So, the more the line is curved, the more the inclusion of the redundant segment (7.4) is localized.

Substituting $\mathbf{r}(l,t)$ to (2.3), we find that the parameter $\tau$ has the meaning of the curve's torsion.

According to (7.3) the soliton moves steadily along the vortex filament with the velocity

$$\upsilon = 2\nu\tau \qquad (7.6)$$

The curve rotates around the $x$ axis with the angular velocity

$$\omega = \frac{\partial(y+iz)/\partial t}{(y^2+z^2)^{1/2}}$$

We may rewrite (7.2) in a more convenient form

$$(\hat{\kappa}-1/a)^2 + \tau^2 = 1/a^2$$

Now, assuming that $a$ is constant, it is easily seen (Fig.4) how the longitudinal extension of the disturbance, measured by $1/\hat{\kappa}$, affects the curve's torsion and the corollaries.

When the disturbance is most localized i.e. the curvature is maximal $\hat{\kappa} = 2/a$, then $\tau = 0$ (Fig.4). That is, the curve is plane. It has the form of a loop. In accord with (7.6) the plane loop is translationally at rest. It rotates steadily around the $x$ axis with the angular velocity $\omega = \nu\hat{\kappa}^2$.

As the disturbance's spread increases from $a/2$ to $a$ i.e. as the curvature decreases to $\hat{\kappa} = 1/a$, the torsion $\tau$ grows to its maximal value $1/a$ (Fig.4). It corresponds to maximal value of the soliton's translational velocity (7.6):

$$\upsilon \leq 2\nu/a \qquad (7.7)$$

In this range of the delocalization

$$\tau/\hat{\kappa} < 1$$

The filament is convolved into the loop (Fig.5, bottom) and the direction of its rotation coincides with that of the vorticity of the unperturbed filament.

Further, as the disturbance delocalizes from $a$ to $\infty$, the torsion drops from $1/a$ to the zero (Fig.4). In this range

$$\tau/\hat{\kappa} > 1$$

The loop is unfolded (Fig.5, top) and thus the rotation becomes opposite to the vorticity.

When $\hat{\kappa} \to 0$, we have $\tau \to 0$. In this event

$$\hat{\kappa} \ll \tau \qquad (7.8)$$

and, in accord with (4.7), the curve tends to an asymptotic helix. This is the humped helix approximated by the wave packet (6.1). The asymptotic helix rotates steadily around the $x$ axis with the angular velocity (4.9) $\omega = \nu\tau^2$.

Differentiating (2.5) with respect to $l$ and using formula (2.1), we get a positionally invariant form of the motion law

$$\frac{\partial \mathbf{e}}{\partial t} = \nu\,\mathbf{e} \times \frac{\partial^2 \mathbf{e}}{\partial l^2}$$

It was shown rigorously [7] that with (2.2), (2.3) this equation can be transformed to the nonlinear Schroedinger equation

$$-\frac{i}{\nu}\frac{\partial \Phi}{\partial t} = \frac{\partial^2 \Phi}{\partial l^2} + \tfrac{1}{2}|\Phi|^2\Phi \qquad (7.9)$$

under the substitution

$$\Phi = \kappa \exp\left[i\left(\int_0^l \tau\, dl - \omega t\right)\right] \qquad (7.10)$$

where $\omega = \text{const}$ is the energy integral of motion.

When

$$\kappa \ll \tau$$

the second term in the right-hand side of (7.9) can be neglected and the equation linearized to (5.2). In this event

$$\Phi \to \varphi = \kappa \exp\left[i(\tau x - \nu \tau^2 t)\right] \qquad (7.11)$$

where $\kappa \to a\tau^2$. So, the wave function takes the meaning of the helix rotation velocity (4.8).

## 8. Integrals of motion

With (7.10) equation (7.9) can be presented in the quasihydrodynamic form

$$\frac{\partial \rho}{\partial t} + \frac{\partial(\rho \upsilon)}{\partial l} = 0 \qquad (8.1)$$

$$\frac{\partial(\rho \upsilon)}{\partial t} + \frac{\partial}{\partial l}\left[\rho \upsilon^2 - \nu^2 \rho \frac{\partial^2 \ln \rho}{\partial l^2} - \tfrac{1}{2}\nu^2 \rho^2\right] = 0 \qquad (8.2)$$

where

$$\rho = \kappa^2 = |\Phi|^2 \qquad (8.3)$$

$$\upsilon = 2\nu\tau \qquad (8.4)$$

Through (2.4), the part $\varepsilon$ of the kinetic energy of the fluid due to distortion of the vortex filament is given by

$$\varepsilon = \tfrac{1}{2}\varsigma \int u^2 dl = \tfrac{1}{2}\varsigma \nu^2 \int \kappa^2 dl = \tfrac{1}{2}\varsigma \nu^2 \int \rho\, dl \qquad (8.5)$$

where $\varsigma$ stands for linear density of the fluid along the filament and (8.3) was used. The energy $\varepsilon$ has the meaning of the self-energy of the disturbance and can be interpreted as the mass $m_\varepsilon$ of this disturbance. By virtue of the continuity equation (8.1), this quantity is conserved:

$$\frac{\partial}{\partial t}\int \rho\, dl = 0$$

Thus, the density of the distribution of the distortion energy along the vortex filament corresponds to the linear density of the space distribution of the soliton's mass $m_\varepsilon$:

$$m_\varepsilon \frac{\tfrac{1}{2}\varsigma u^2}{\varepsilon} = m_\varepsilon \frac{\tfrac{1}{2}\varsigma \nu^2 \rho}{\varepsilon} \qquad (8.6)$$

In these terms, the flow of the distortion energy along the filament

$$\tfrac{1}{2}\varsigma \nu^2 \rho \upsilon$$

acquires the meaning of the soliton's local momentum. From the dynamic equation (8.2) you see that the total momentum of the soliton is conserved:

$$\frac{\partial}{\partial t}\int \rho \upsilon\, dl = 0 \qquad (8.7)$$

Next, using (8.6), the soliton's translational energy can be considered

$$E_t = \tfrac{1}{2}\int m_\varepsilon \frac{\tfrac{1}{2}\varsigma \nu^2 \rho}{\varepsilon} \upsilon^2 dl \qquad (8.8)$$

The contribution of the diffusion flow

$$\rho w = -v \frac{\partial \rho}{\partial l}$$

should be also taken into account. In the nonlinear case, we must include into the integral the term from (8.2) of the binding energy

$$-\tfrac{1}{2} v^2 \rho^2$$

Then the total energy is conserved:

$$\frac{\partial}{\partial t} \int \tfrac{1}{2} \rho (v^2 + w^2 - v^2 \rho) \, dl = 0$$

We may also add to the dynamic equation (8.2) the density of the external force. For the potential force it looks as

$$\frac{\partial (\rho v)}{\partial t} + \frac{\partial}{\partial l} \left[ \rho v^2 - v^2 \rho \frac{\partial^2 \ln \rho}{\partial l^2} - \tfrac{1}{2} v^2 \rho^2 \right] + \rho \frac{\partial U}{\partial l} = 0$$

When the potential $U(l)$ does not depend on the time, the following quantity is conserved

$$\int \rho \left[ \tfrac{1}{2} (v^2 + w^2 - v^2 \rho) + U \right] dl$$

Thence, the nonlinear Schroedinger equation with the potential energy $U$ should be written as

$$-\frac{i}{v} \frac{\partial \Phi}{\partial t} = \frac{\partial^2 \Phi}{\partial l^2} + \tfrac{1}{2} |\Phi|^2 \Phi - \frac{1}{v^2} U \Phi \qquad (8.9)$$

Substituting (7.5) to the second integral in (8.5), we get the soliton's self-energy:

$$\varepsilon = 4 \varsigma v^2 \hat{\kappa} \qquad (8.10)$$

The soliton's translational energy $E_t$ is easily found substituting (8.4) with $\tau = \text{const}$ to (8.8):

$$E_t = m_\varepsilon \frac{2 v^2 \tau^2}{\varepsilon} \tfrac{1}{2} \varsigma v^2 \int \rho \, dl$$

Then, using in this expression (8.5), we get

$$E_t = 2 m_\varepsilon v^2 \tau^2 \qquad (8.11)$$

In asymptotics, when $\hat{\kappa} / \tau \to 0$, relation (7.2) reduces itself to

$$2 \hat{\kappa} = a \tau^2 \qquad (8.12)$$

Then we have for (8.10)

$$\varepsilon = 2 \varsigma v^2 a \tau^2 \qquad (8.13)$$

You see that the expression (8.13) for the fluid energy coincides with that (8.11) for the soliton's kinetic energy if we take for the mass of the asymptotic soliton

$$m_\varepsilon = \varsigma a \qquad (8.14)$$

That enables us to identify the energy integral of motion of the asymptotic soliton with the real energy of the fluid motion, and the mass of the soliton – with the real mass (8.14) of the fluid.

## 9. Particle

In this section we will demonstrate with a simplified model that there exists a singular unique size of the loop-shaped soliton on a vortex filament. The redundant segment (7.4) of the filament, needed in order to form the curvilinear configuration on the originally straight line, brings with itself the energy of the fluid motion

$$2 a \xi \qquad (9.1)$$

where $\xi$ is the energy density on a unit length of the filament. The energy of distortion associated with the loop is given by (8.10). For the plane configuration of the loop it equals to

$$\varepsilon = \frac{8\varsigma v^2}{a} \qquad (9.2)$$

where (7.2) with $\tau = 0$ was used. Summing (9.1) and (9.2), we find the total energy of the fluid associated with the loop as a function of the redundant length $2a$:

$$2a\xi + \frac{8\varsigma v^2}{a} \qquad (9.3)$$

This function has a minimum at

$$a = 2v\left(\frac{\varsigma}{\xi}\right)^{1/2}$$

which determines the singular size of the loop. The same is valid with respect to the vortex ring obtained from the loop by reconnection of the filament at the point of intersection.

For visuality, let us reproduce the whole argumentation for the vortex ring. The fluid energy associated with the length is evaluated by

$$2\pi R \xi$$

where $R$ is the radius of the ring. Its curvature is given by

$$\kappa = \frac{1}{R}$$

So, the energy of distortion associated with the ring can be found from the integral in (8.5) as

$$\varepsilon = \tfrac{1}{2}\varsigma v^2 \left(\frac{1}{R}\right)^2 2\pi R$$

Then the total energy is given by

$$2\pi \xi R + \frac{\pi \varsigma v^2}{R}$$

Comparing it with (9.3), wee see that $a$ has the meaning of the loop's diameter.

The filament is taken in the current model as the idealization of the vortex tube. In the perfect fluid the vortex tube is hollow inside. So, the curvilinear configuration of the tube – the helix, the loop or the vortex ring – just corresponds to the inclusion of a redundant void in the discrete structure of the vortex sponge. That agrees well with the mesoscopic mechanical model of a particle [3, 4]. Although, the mass $\varsigma 2a$ of the redundant segment (7.4) of the vortex tube appears to be twice the mass of the disturbance that is computed using the formula (8.14) for asymptotic helix.

As you see, provided that the strength of the vortex tube is fixed, the construction described ensures the discreteness of a nonlinear configuration in the structure of the vortex sponge.

A plane loop on a vortex filament can't be split into smaller plane loops without the input of some fluid energy. Indeed, let it be divided into two parts $\alpha$ and $1-\alpha$, where $1 > \alpha > 0$. As you see from (9.1), (7.4) the energy of the background is additive and thus does not change in splitting. Whereas the energy of disturbance computed with (9.2) will increase:

$$\frac{1}{\alpha} + \frac{1}{1-\alpha} > 1$$

However, the plane loop can be split into non-planar solitons i.e. into waves. This process needs some increase in the secondary integral of energy (8.8). Thus, we have from (7.2) that the plane loop with the curvature $\kappa$ (7.5) can be split into $m$ waves with the curvature $\kappa/m$ and for $m \gg 1$ with the torsion $\tau \approx \hat{\kappa}$. According to (7.6), a non-planar soliton moves translationally with the velocity $2v\tau$. Requiring the conservation (8.7) of the momentum, we find that the splinters will move in opposite directions.

In asymptotics, when $\kappa/\tau \to 0$, we have (8.13) instead of (9.2). Now the distortion energy is additive with respect to division of the redundant segment (7.4). So, the helix can be split as a classical mass body.

## 10. Elementary helix

As you see in (7.7), the velocity of the given soliton is restricted from the above by

$$\upsilon_{max} = 2\nu/a$$

(Fig.4). In its turn, $\upsilon_{max}$ is restricted by some fundamental constant $c$, which must be the speed of the perturbation wave in the turbulent medium:

$$\upsilon_{max} \leq c$$

That implies

$$a \geq 2\nu/c$$

Thus we come to the concept of the elementary inclusion, having the minimal size of the redundant segment

$$a_0 = 2\nu/c$$

It probably exists only as an asymptotic helix. In this model $c$ corresponds [2] to the speed of light in vacuum.

On the other side, the condition (7.8) of the asymptotic case can be written as

$$\frac{2\hat{\kappa}}{\tau} \leq \beta \ll 1$$

where $\beta$ is an upper bound for the asymptoticity. Combining it with (8.12), (7.6) gives

$$\upsilon \leq \frac{2\nu}{a}\beta$$

That shows that the domain of velocities, for which the linear Schroedinger equation is valid, broadens with the decrease in the length of the redundant segment (Fig.4, the left side).

So, in order to increase the maximal velocity of the disturbance, we must divide the inclusion of the redundant segment $2a$ into parts.

## 11. Thermalization

Supposedly, under the action of the stochastic medium the soliton on a vortex filament splits into the elementary helices above mentioned.

We see the thermalized soliton as a system of $m$ identical segments $a_0 = a/m$ each of which obeys the linear Schroedinger equation (5.2). From it a single many-body equation can be formally composed

$$\frac{\partial \Psi}{\partial t} = i\nu \sum_{n=1}^{m} \frac{\partial^2 \Psi}{\partial x_n^2} \qquad (11.1)$$

where the function $\Psi$ is given by the product of the forms (7.11)

$$\Psi = \prod_{n=1}^{m} \kappa_n \exp\left[i\left(\tau_n x_n - \nu\tau_n^2 t\right)\right] \qquad (11.2)$$

Passing in (11.1) to the center point variable

$$x = \frac{1}{m}\sum_{n=1}^{m} x_n \qquad (11.3)$$

we may get via the well-known procedure the equation

$$\frac{\partial \psi}{\partial t} = i\frac{\nu}{m}\frac{\partial^2 \psi}{\partial x^2} \qquad (11.4)$$

This rather a formal result can be visualized if we consider the phase of the wave function (11.2), which is taken for $m$ helices with equal values $\tau_n = \tau$ of the torsion:

$$\sum_{n=1}^{m}\left(\tau_n x_n - v\tau_n^2 t\right) \to \tau\sum_{n=1}^{m} x_n - mv\tau^2 t = (m\tau)x - \frac{v}{m}(m\tau)^2 t = kx - \frac{v}{m}k^2 t \qquad (11.5)$$

where

$$k = m\tau \qquad (11.6)$$

and $x$ is given by (11.3).

Provided that the length $2a_0$ of an elementary segment is constant, the number $m$ of elementary constituents involved in soliton can be taken as a measure of the soliton's mass, the real mass being

$$m_\varepsilon = \varsigma a = m\varsigma a_0$$

where (8.14) was used. Then the quantity $k$ above defined (11.6) can be taken as a measure of the soliton's momentum

$$p = \varsigma a\upsilon = \varsigma a_0\, 2vm\tau = 2v\varsigma a_0 k$$

where (8.4) was used. The frequency term in the phase (11.5) of the wave function acquires the meaning of the soliton's kinetic energy (8.11)

$$E = \frac{p^2}{2m\varsigma a_0} = 2v\varsigma a_0 \frac{vk^2}{m}$$

In quantum mechanics the constant analogous to $v$ is usually designated as

$$v\varsigma a_0 = \frac{\hbar}{2}$$

## 12. Collapse

A fluctuation of the fluid pressure may cause the spatial distribution of splinters to re-collect into the original soliton. We describe this process phenomenologically adding to (11.1) the pairwise attraction between the elementary helices

$$\frac{\partial \Psi}{\partial t} = iv\left(\sum_{n=1}^{m}\frac{\partial^2 \Psi}{\partial x_n^2} - \frac{1}{v^2}U\Psi\right) \qquad (12.1)$$

where the mass density of the potential

$$U = -\frac{4v^2\hat{\kappa}}{m}\sum_{s<q}^{m}\delta(x_s - x_q) \qquad (12.2)$$

In (12.1) the potential was introduced in the same way as it was done in (8.9). In (12.2) the coefficient before the $\delta$-function was chosen in accord with (8.10) assuming that the self-energy of the fragment is $1/m$ of the self-energy (8.10) of the original soliton.

Equation (12.1) with (12.2) can be resolved exactly. However, it will be illuminative to give the scheme based on Hartree approximation

$$\Psi = \prod_{n=1}^{m}\varphi_n(x_n, t) \qquad (12.3)$$

where $\varphi_n$ is the wave function of a splinter:

$$\frac{\partial \varphi_n}{\partial t} = iv\frac{\partial^2 \varphi_n}{\partial x_n^2}$$

Compare (12.3) with (11.2). By (8.5), (8.3), the self-energy of the splinter is computed via

$$\tfrac{1}{2}\varsigma v^2\int|\varphi_n|^2 dx_n$$

It was taken in (12.2) to be $1/m$ of the self-energy (8.10) of the original soliton:

$$4\varsigma v^2 \hat{\kappa}/m$$

That implies the following normalization of $\varphi_n$

$$\int |\varphi_n|^2 dx_n = 8\hat{\kappa}/m \qquad (12.4)$$

$$n = 1,2,...m$$

Substituting (12.3) to (12.1) with (12.2), multiplying it by

$$\prod_{n=2}^{m} \varphi_n^*(x_n,t) \, dx_n$$

and then integrating over all $x_n$, where $n \neq 1$, we get

$$-\frac{i}{v}\frac{\partial \varphi}{\partial t} = \frac{\partial^2 \varphi}{\partial x^2} + \tfrac{1}{2}(m-1)|\varphi|^2 \varphi + \frac{m(m-1)(m-2)}{32\hat{\kappa}} \varphi \int |\varphi|^4 dx$$

Taking

$$\phi = (m-1)^{1/2} \varphi \exp(-i\omega_o t) \qquad (12.5)$$

where

$$\omega_0 = v \frac{m(m-1)(m-2)}{32\hat{\kappa}} \int |\varphi|^4 dx$$

we come to

$$-\frac{i}{v}\frac{\partial \phi}{\partial t} = \frac{\partial^2 \phi}{\partial x^2} + \tfrac{1}{2}|\phi|^2 \phi$$

This equation coincides with (7.9) when $l \to x$ i.e. when $\kappa << \tau$. Take notice that for $m >> 1$ (12.5) with (12.4) gives the following normalization of the wave function $\phi$:

$$\tfrac{1}{2}\int |\phi|^2 dx = 4\hat{\kappa}$$

Compare it with (8.10) obtained from (8.5), (8.3) with (7.5)

$$\varepsilon = \tfrac{1}{2}\varsigma v^2 \int |\Phi|^2 dl = 4\varsigma v^2 \hat{\kappa}$$

So, the quantum definition of the particle's mass given in section 11 agrees with its mechanical definition given in section 8.

Similar results can be obtained in a simpler model: if we take in (12.2) $s = 1$, $q = 2,..., m$.

The choice of the place or splinter, where the soliton will be re-collected, is the competence of a more general model of the measurement, which the above scheme should be included in. One may expect that it is probabilistic and the probability density is proportional to local decrease in the fluid pressure. By hydrodynamics [3] the decrement of the pressure equals to the increase in the energy density i.e. it is proportional to $|\psi|^2$ from (11.4) or similar equation.

## 13. Spin

There are two kinds of the helix differed in the sign of the torsion $\tau$. The right-hand screw helix (Fig.2, bottom) is described by (4.1)

$$\begin{aligned} y &= a\cos(\tau x) \\ z &= a\sin(\tau x) \end{aligned} \qquad (13.1)$$

with $\tau > 0$. The left-hand screw one (Fig.2, top) is described by (13.1) with $\tau < 0$. According to (4.8), the helix rotates around the $x$ axis with the velocity

$$\mathbf{u} = a\nu\tau^2[\sin(\tau x)\,\mathbf{i}_2 - \cos(\tau x)\,\mathbf{i}_3] \qquad (13.2)$$

As you can see from it, both kinds rotate in one and the same direction - counter to the direction of the filament's vorticity, which in (13.2) was chosen such as to coincide with direction of the $x$ axis.

According to (7.6), or (6.2), the helix moves translationally with the velocity

$$\upsilon = 2\nu\tau$$

That is, the right-hand screw helix travels in the direction, which the filament's vorticity points to. While the left-hand screw helix goes in the opposite direction.

In three dimensions we deal with the ideal fluid pierced in all directions by the vortex filaments [5]. Macroscopically (to be fine, mesoscopically) this system looks as a turbulent ideal fluid. Perturbations of the turbulence was shown [2] to reproduce the electromagnetic fields. In particular, the average fluid velocity $\langle\mathbf{u}\rangle$ corresponds to the magnetic vector-potential. The rotation of the soliton is seen macroscopically as a singularity – the center of torsion in the quasielastic medium. It corresponds to a magnetic dipole $\boldsymbol{\mu}$. The energy of its interaction with the external vorticity field is given by

$$-\boldsymbol{\mu}\cdot\operatorname{curl}\langle\mathbf{u}\rangle \qquad (13.3)$$

The fluid vorticity $\operatorname{curl}\langle\mathbf{u}\rangle$ just corresponds to the magnetic field.

Let two kinds of the helices move in the turbulent substratum from the left to the right. The first helix (Fig.6, bottom) is right-hand screw and hence it moves along a filament whose vorticity is also directed to the right. The other helix (Fig.6, top) is left-hand screw. So, it moves to the right along a filament whose vorticity is directed opposite to direction of the motion. The question is how an observer may distinguish between these two cases?

It can be done imposing on them the external field of fluid vorticity $\operatorname{curl}\langle\mathbf{u}\rangle$. So, we have the conditions of the Stern-Gerlach experiment. The vertical arrow at Fig.6 just shows the fluid vorticity directed and growing from the bottom to the top. This inhomogeneous vorticity field will deflect the traveling helices such as to diminish their energy in accord with formula (13.3). In order to change somewhat the direction of its motion, a helix must jump over to the adjacent filament with similar but slightly different direction of vorticity. Thus, the helices will behave themselves in the same way (see Fig.6) as spin particles in the real Stern-Gerlach experiment.

## 14. Concluding remark

The mechanical model above constructed reproduces the main features of a microparticle including its discreteness. However, there is a point that should be still elucidated. This is the discrete structure of the vortex sponge i.e. the fixed strength of the intrinsic vortex tube, or filament. At the time being, it is taken as a postulate.

**References**

[1] V.P.Dmitriyev, " Particles and charges in the vortex sponge", *Z.Naturforsch.* **48a** No 8/9 (1993) 935-942.

[2] O.V.Troshkin, " On wave properties of an incompressible turbulent fluid", *Physica A* **168** No 2 (1990) 881-899.

[3] V.P.Dmitriyev, "Towards an exact mechanical analogy of particles and fields", *Nuovo Cimento* **111A** No 5 (1998) 501-511; http://xxx.lanl.gov/abs/physics/9904029

[4] V.P.Dmitriyev, "Mechanical analogies for the Lorentz gauge, particles and antiparticles", *Apeiron* **7** No 3/4 (2000) 173-183; http://xxx.lanl.gov/abs/physics/9904049

[5] E.M.Kelly, "Vacuum electromagnetics derived exclusively from the properties of an ideal fluid", *Nuovo Cimento* **32B** No 1 (1976) 117-137.

[6] G.K.Batchelor, *An introduction to fluid dynamics*, University Press, Cambridge, 1970.

[7] H.Hasimoto, "A soliton on a vortex filament", *J. Fluid Mech.* **51** part 3 (1972) 477-485.

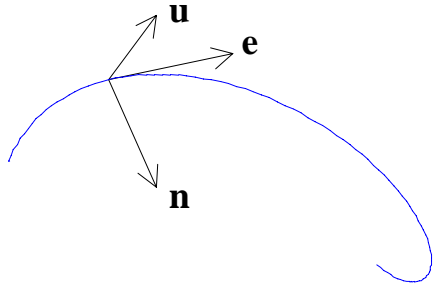

Fig. 1. The drift **u** of a bent vortex filament in relation to its curvature $\kappa\mathbf{n}$ and vorticity **e**.

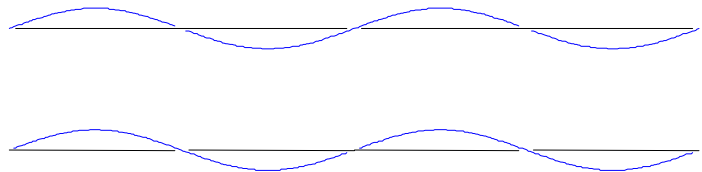

Fig. 2. The right-hand screw helix (bottom) and the left-screw helix (top) in relation to the $x$ axis. The $xz$ projection of (4.1) or (13.1) is shown.

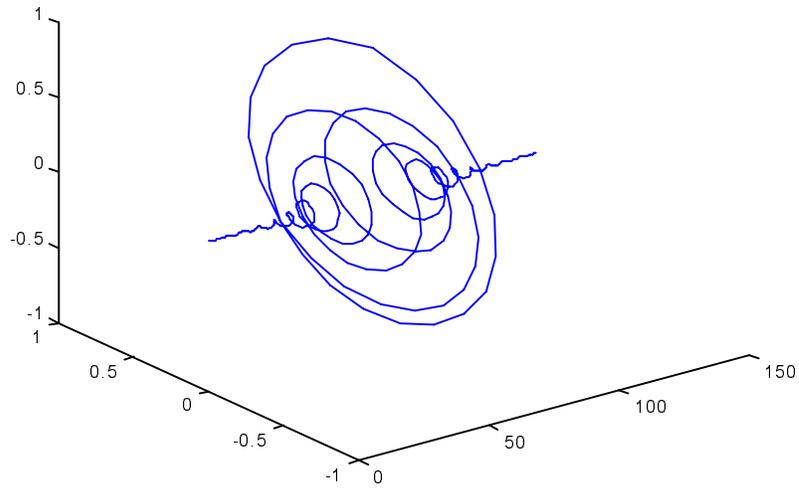

Fig. 3. A wave packet (6.1).

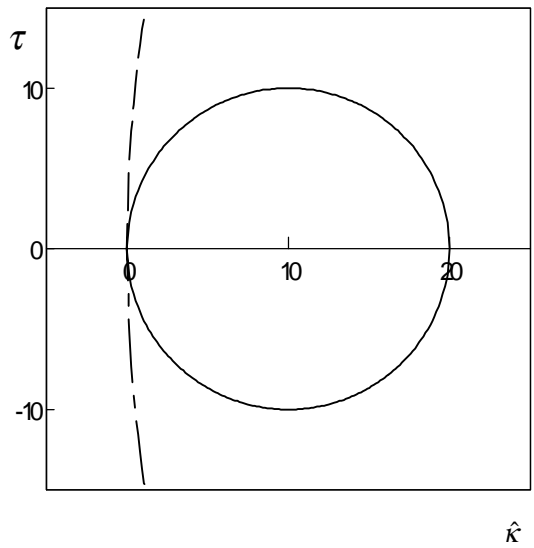

Fig. 4. The torsion $\tau$ in relation to curvature parameter $\hat{\kappa}$ plotted by (7.2) with $a = $ const. Solid line: $a = 0.1$; dashed line: $a = 0.01$.

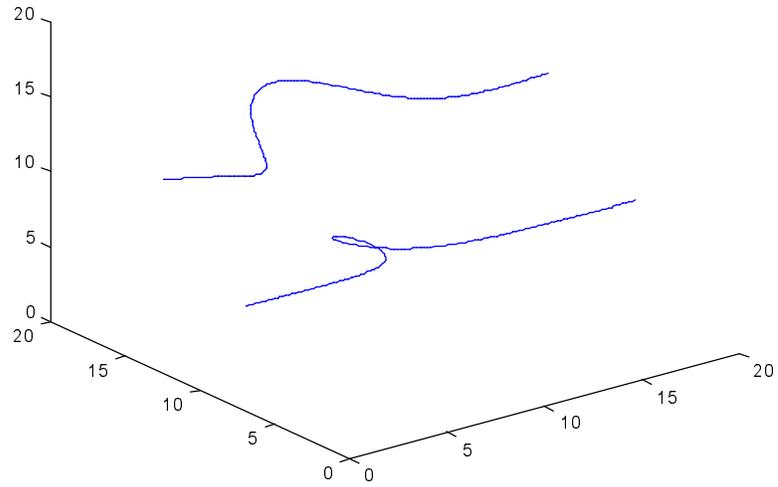

Fig. 5. The loop-shaped soliton on a vortex filament at $\tau/\hat{\kappa}=0.23$ (bottom) and $\tau/\hat{\kappa}=1.1$ (top).

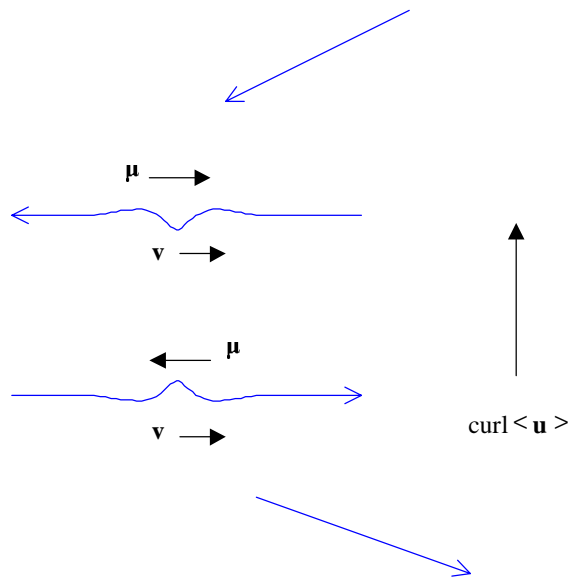

Fig. 6. A right-hand screw helix (bottom) and left-hand screw helix (top) traveling from the left to the right in the vortex sponge through an inhomogeneous field of fluid vorticity curl $\langle\mathbf{u}\rangle$. Arrows on the filaments indicate the direction of their vorticity, $\mathbf{v}$ shows the direction of the translational motion and $\mathbf{\mu}$ the rotational moment of the helices.